\documentclass[conference]{IEEEtran}
\IEEEoverridecommandlockouts
\usepackage{cite}
\usepackage{amsmath,amssymb,amsfonts}
\usepackage{algorithmic}
\usepackage{graphicx}
\usepackage{textcomp}
\usepackage{xcolor}
\usepackage{subcaption}
\usepackage{multirow}
\usepackage{multicol}
\usepackage{array}

\def\BibTeX{{\rm B\kern-.05em{\sc i\kern-.025em b}\kern-.08em
    T\kern-.1667em\lower.7ex\hbox{E}\kern-.125emX}}
\begin{document}

\title{Design of a Low-Power High-Gain Bio-Medical Operational Amplifier in 65nm Technology using gm/ID Methodology\\
}

\author{\IEEEauthorblockN{1\textsuperscript{st} Ayan Biswas}
\IEEEauthorblockA{\textit{Dept. of Electronics and Tele-communication Engineering} \\
\textit{Jadavpur University}\\
Kolkata – 700 032, INDIA \\
ayanbiswas@ieee.org}
\and
\IEEEauthorblockN{2\textsuperscript{nd} Supriya Dhabal}
\IEEEauthorblockA{\textit{Electronics \& Communication Engineering Department} \\
\textit{Netaji Subhash Engineering College}\\
Kolkata - 700152, INDIA \\
supriya\_dhabal@yahoo.co.in }
\and
\IEEEauthorblockN{3\textsuperscript{rd} Palaniandavar Venkateswaran}
\IEEEauthorblockA{\textit{Dept. of Electronics and Tele-communication Engineering} \\
\textit{Jadavpur University}\\
Kolkata – 700 032, INDIA \\
pvwn@ieee.org}
}

\maketitle

\begin{abstract}
Operational Amplifiers (Op-Amps) play a crucial role in the field of biomedical engineering, as they enable signal amplification and processing in various medical devices. With the increasing demand for portable and low-power biomedical devices, designing Op-Amps specifically tailored for such applications is essential. In response to this need, a low-power high-gain Op-Amp designed for biomedical applications using TSMC 65nm technology has been proposed. This Op-Amp incorporates a two-stage miller compensated topology, which is well-known for its superior performance in gain, gain bandwidth product and power consumption. The proposed Op-Amp contributes to the field of biomedical engineering by offering a tailored solution that enhances signal processing capabilities, enables accurate data acquisition, and improves overall efficiency in healthcare systems. The design methodology and simulation results presented in this paper provide insights into the performance and potential impact of the Op-Amp in advancing biomedical devices and systems.

    \begin{figure}[ht]
        \centering
        \includegraphics[width=0.85\linewidth]{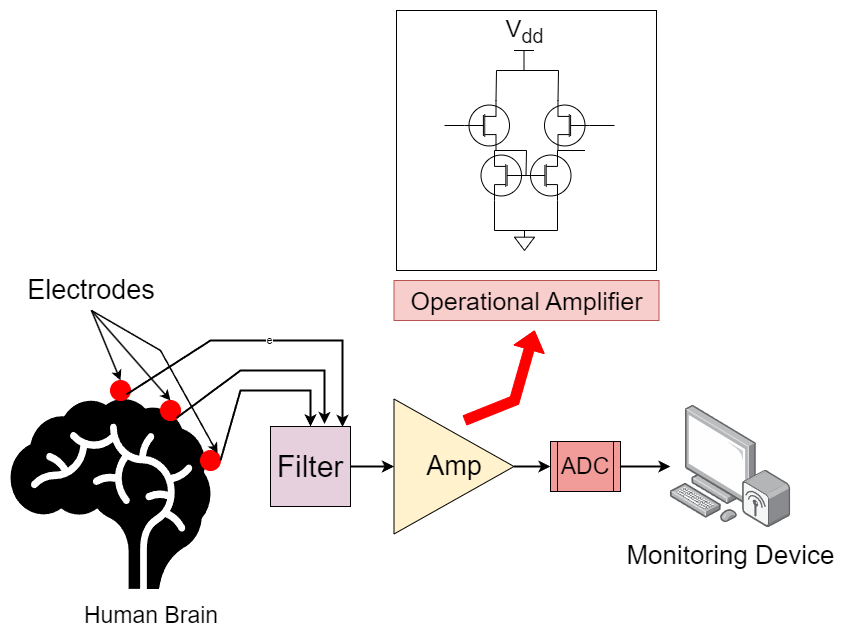}
        \caption{Design of a common EEG signal monitoring system}
        \label{fig:pic-abs}
    \end{figure}
    
\end{abstract}


\section{Introduction}

The miniaturization of VLSI technologies demands low-voltage and low-power analog circuit designs~\cite{Rami}. However, reduced supply voltage presents challenges for analog circuit designers due to degraded transistor characteristics and limitations in traditional techniques. This is crucial for bio-measurement devices, such as portable and battery-powered devices. Bio-potential amplifiers are commonly used to amplify ECG and EEG signals, requiring high input impedance, safety protections, low output impedance, minimal distortion, high gain, and a high common mode rejection ratio. Precise measurement and amplification of small and noisy bio-signals pose challenges~\cite{Vullings}.

This work aims to develop an operational amplifier for improved biomedical signal analysis (e.g. EEG and ECG signals). The objective is to design a low-noise, low-power amplifier to enhance measurement accuracy and reliability. By addressing challenges in signal analysis, our innovative design aims to improve monitoring quality and efficiency for medical professionals and patients.



\section{Design Methodology}
This section outlines a step-by-step process for creating a two-stage operational amplifier (also known as a Miller amplifier), including specific dimensions for the MOSFETs (including device width (W), and channel length (L)), and the required value for the compensation capacitor ($C_{c}$) to meet the desired performance of the amplifier

\subsection{Choice of Devices}

The chosen devices for the design are the \textbf{nch\_lvt\_mac} and \textbf{pch\_lvt\_mac} from the \textbf{tsmcN65} library. The \textbf{nch\_lvt\_mac} is an n-type MOSFET with low threshold voltage and a metal-gate structure, suitable for low-power and high-performance applications. The \textbf{pch\_lvt\_mac} is a complementary p-type MOSFET with similar characteristics, making it suitable for amplifier circuits.

\subsection{The gm/id methodology}

Our Op-Amp design utilizes the gm/ID methodology, simplifying the process with pre-generated sizing charts (shown in Fig.~\ref{fig:sizing-pmos} and Fig.~\ref{fig:sizing-nmos}) instead of complex equations. This approach achieves desired specifications in a single iteration, capturing MOSFET behavior across various inversion regions. Inspired by Hesham et al.~\cite{Hesham}, who applied similar techniques to FinFET devices, our methodology enables specification-driven design without relying on compact models.

\subsection{Bias Circuitry for generation of design variable curves}

\begin{figure*}[h]
    \centering
    \begin{minipage}{0.45\textwidth}
        \centering
        \includegraphics[width=0.85\linewidth, height = 150pt]{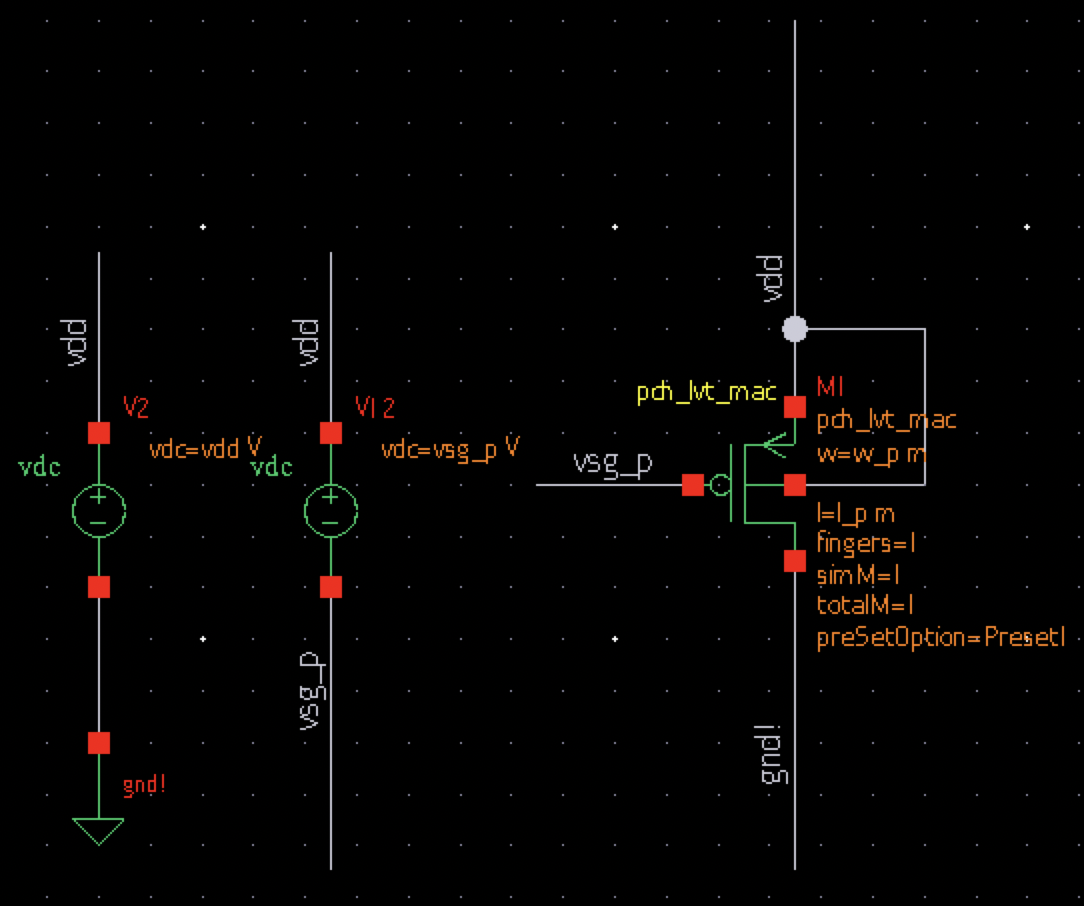}
        \caption{PMOS bias circuitry for sizing chart generation}
        \label{fig:schematic-pmos-bias}
    \end{minipage}
    \hfill
    \begin{minipage}{0.45\textwidth}
        \centering
        \includegraphics[width=0.85\linewidth, height = 150pt]{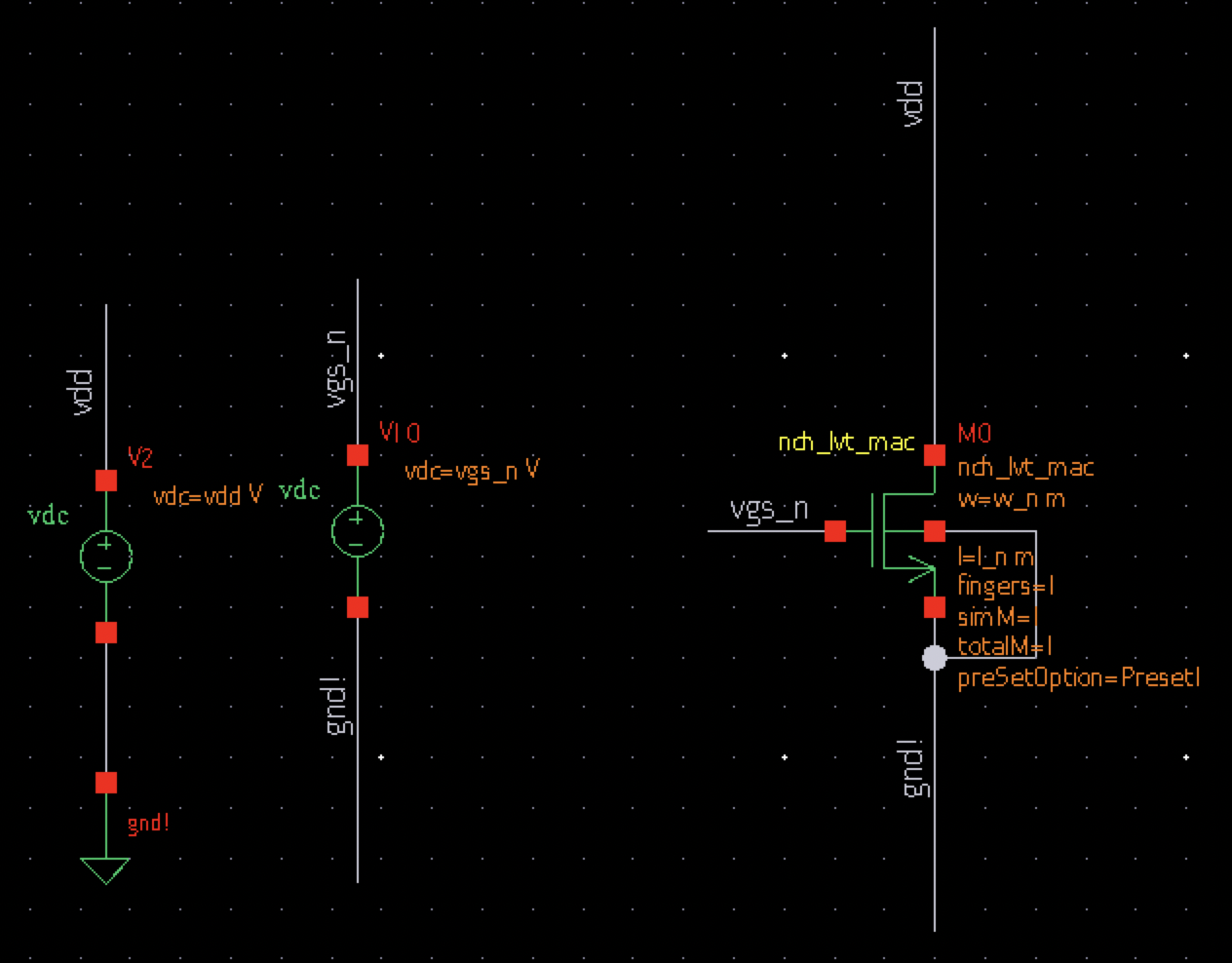}
        \caption{NMOS bias circuitry for sizing chart generation}
        \label{fig:schematic-nmos-bias}
    \end{minipage}
\end{figure*}

Fig.\ref{fig:schematic-pmos-bias} depicts the PMOS bias circuitry for generating sizing charts, such as Vov vs gm/id and gm/id vs gm/gds. The NMOS bias circuitry is illustrated in Fig.\ref{fig:schematic-nmos-bias}.

At first, the gm/id curve for both of the devices is plotted against $V_{gs}$ and then by analysis of data from the results of this plot, other plots are generated.

\subsection{Sizing of pMOS and nNMOS Transistors}
\begin{figure*}
    \centering
    \begin{subfigure}{0.32\linewidth}
        \centering
        \includegraphics[width=\linewidth, height=180pt]{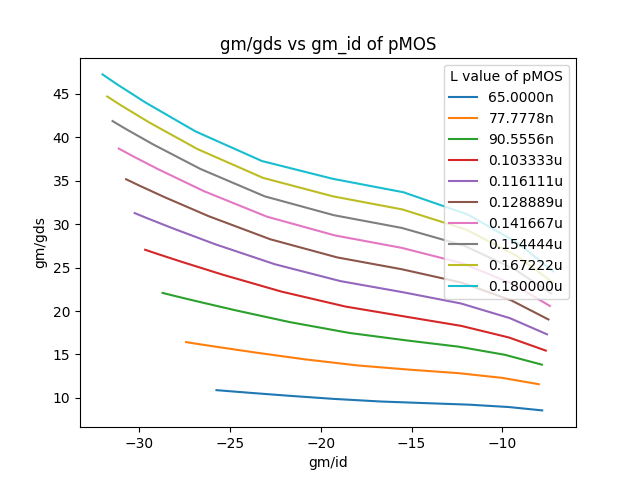}
        \caption{gm/gds vs gm/id curve for \textbf{pch\_lvt\_mac}}
        \label{fig:pmos_gmgd_gmid}
    \end{subfigure}
    \hfill
    \begin{subfigure}{0.32\linewidth}
        \centering
        \includegraphics[width=\linewidth, height=180pt]{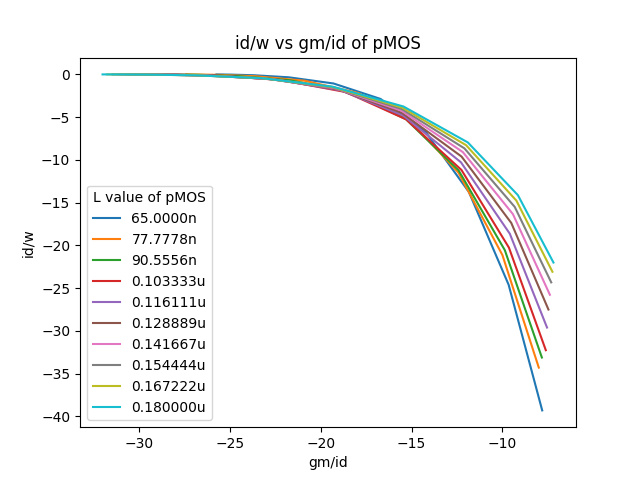}
        \caption{id/w vs gm/id curve for \textbf{pch\_lvt\_mac}}
        \label{fig:pmos_idw_gmid}
    \end{subfigure}
    \hfill
    \begin{subfigure}{0.32\linewidth}
        \centering
        \includegraphics[width=\linewidth, height=180pt]{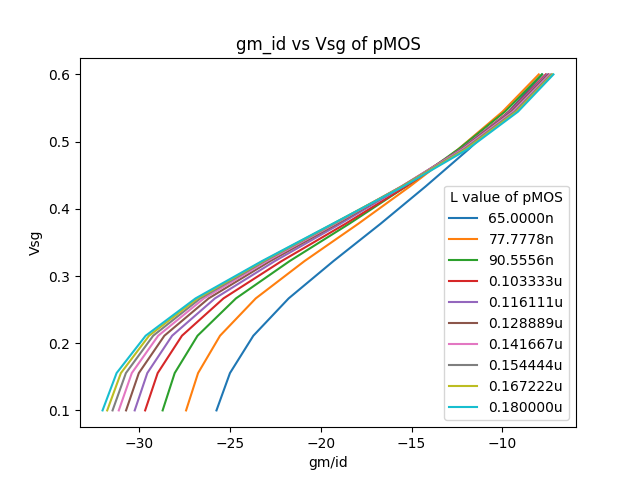}
        \caption{gm/id vs Vsg curve for \textbf{pch\_lvt\_mac}}
        \label{fig:pmos_vsg_gmid}
    \end{subfigure}
    
    \caption{Sizing Charts for pMOS (\textbf{pch\_lvt\_mac}) device where Length (L) varies from 65nm to 180nm}
    \label{fig:sizing-pmos}
\end{figure*}

\begin{figure*}
    \centering
    \begin{subfigure}{0.32\linewidth}
        \centering
        \includegraphics[width=\linewidth, height=180pt]{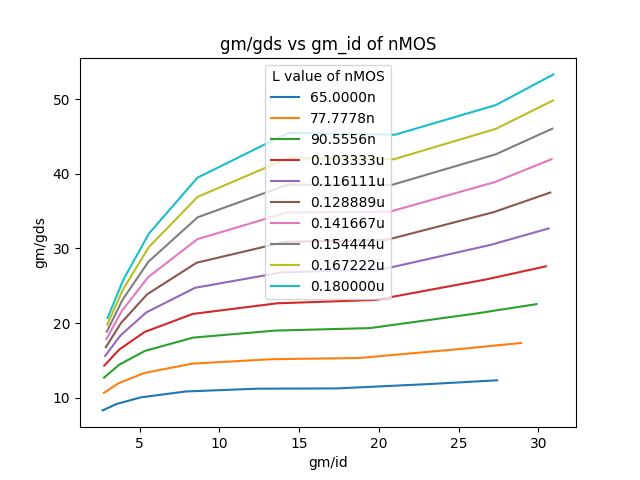}
        \caption{gm/gds vs gm/id curve for \textbf{nch\_lvt\_mac}}
        \label{fig:nmos_gmgd_gmid}
    \end{subfigure}
    \hfill
    \begin{subfigure}{0.32\linewidth}
        \centering
        \includegraphics[width=\linewidth, height=180pt]{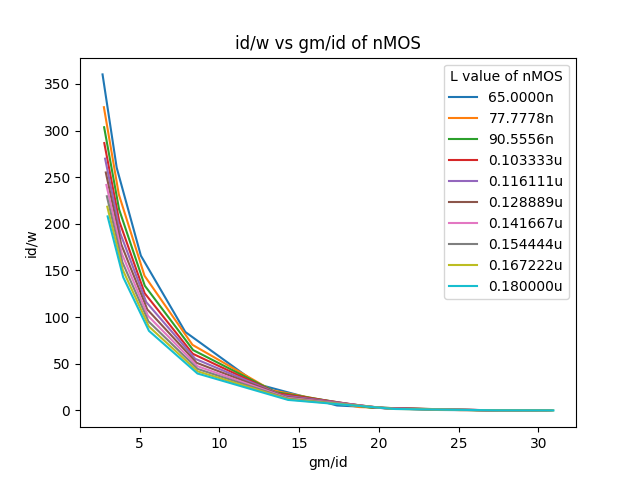}
        \caption{id/w vs gm/id curve for \textbf{nch\_lvt\_mac}}
        \label{fig:nmos_idw_gmid}
    \end{subfigure}
    \hfill
    \begin{subfigure}{0.32\linewidth}
        \centering
        \includegraphics[width=\linewidth, height=180pt]{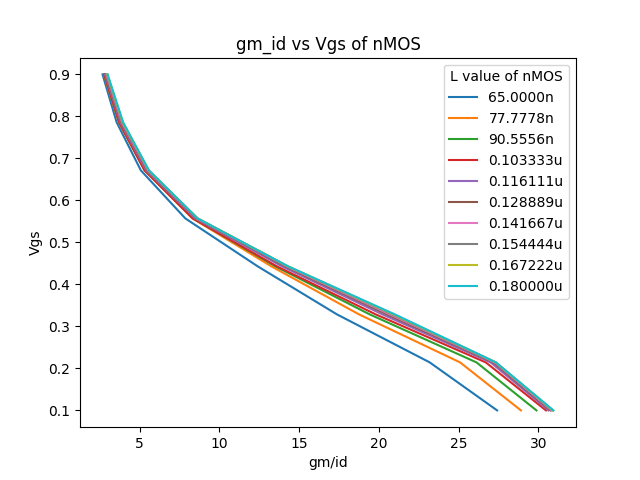}
        \caption{gm/id vs Vgs curve for \textbf{nch\_lvt\_mac}}
        \label{fig:nmos_vgs_gmid}
    \end{subfigure}
 
    \caption{Sizing Charts for nMOS (\textbf{nch\_lvt\_mac}) device where Length (L) varies from 65nm to 180nm}
    \label{fig:sizing-nmos}
\end{figure*}

As shown in Table.~\ref{tb:sizing_sweep} the variables are swept over the desired ranges and corresponding data points have been collected to generate the sizing charts as shown in Fig.~\ref{fig:sizing-pmos} and Fig.~\ref{fig:sizing-nmos}.

\begin{table}[ht]
\centering
\caption{Sweep of Variables for Sizing Chart Generation}
\label{tb:sizing_sweep}
\begin{tabular}{|c|l|l|l|}
\hline
\textbf{MOS Type} & \textbf{Parameter} & \textbf{Sweep Range} & \textbf{Points Taken} \\ \hline
\multirow{2}{*}{pMOS} & Vsg & 0.1 V - 0.9 V & 10 nos \\ \cline{2-4} 
 & L & 65nm - 180nm & 10 nos \\ \hline
\multirow{2}{*}{nMOS} & Vgs & 0.1 V - 0.9 V & 10 nos \\ \cline{2-4} 
 & L & 65nm - 180nm & 10 nos \\ \hline
\end{tabular}
\end{table}

\subsection{Design Equations and Methodology} \label{subsec:method}

The schematic of the two-stage operational amplifier is shown in Fig.~\ref{fig:ckt_mos_level} and the step-by-step design procedure is described below.

\subsubsection*{\textbf{Input Referred Noise Voltage}}
We can start the design from the noise  considerations. At a higher frequency range of operation, the Input Referred Noise Voltage of the amplifier is given by Eq.~\ref{noise-1}~\cite{mahatta}.

\begin{equation} 
    S_{n}(f)=2.4kT\frac{2}{3}\frac{1}{gm_{1,2}}[1+\frac{gm_{3,4}}{gm_{1,2}}] \label{noise-1} 
\end{equation}

For a lower level of noise, we can assume $gm_{3,4} << gm_{1,2} $ so the Eq.~\ref{noise-1} transformed Eq.~\ref{noise-2}

\begin{equation} 
    gm_{1,2}=\frac{16}{3}\frac{kT}{S_{n}(f)} 
    \label{noise-2} 
\end{equation}

So the value of $gm_{1,2} $ can be directly calculated from \ref{noise-2}.

\subsubsection*{\textbf{Miller Capacitance ($C_{c}$)}}

For a given value of the gain-bandwidth product (GBW), the compensation capacitance $C_{c}$ can be calculated from the Eq.~\ref{eq:Cc}~\cite{allen2011cmos}.

\begin{equation} 
    \mathrm{C}_{\mathrm{c}}=\frac{gm_{1,2}}{2\pi GBW_{Hz} }
    \label{eq:Cc}
\end{equation}

The compensation capacitance $C_{c}$ is an essential parameter for tuning the phase margin of the overall design. A phase margin of less than 60 degrees will lead to an unstable signal response at the output.

\subsubsection*{\textbf{Slew Rate}}

Slew Rate (as given by Eq.~\ref{eq:slew_rate}~\cite{mahatta}) is defined as the maximum rate at which an operational amplifier (op amp) can alter its output voltage in response to abrupt changes in the input voltage.

\begin{equation} 
    SR_{I}=\frac{2.I_{D1}}{C_{c}} = \frac{I_{D5}}{C_{c}}
    \label{eq:slew_rate}
\end{equation}

From the given slew rate specification the value of currents i.e. $I_{D5}$ and  $I_{D1}$ can be easily calculated.

\subsubsection*{\textbf{Gain of 1st Stage}}

As we know, Gain is equal to the product of transconductance and output resistance, so here the signal in the 1st stage (as shown in Fig.~\ref{fig:ckt_mos_level}) flows from the input (i.e. M2) and flows to the output (i.e. M4) so transconductance is $gm_{1,2}$ and the overall output resistance is $\frac{1}{gds_{1,2}+gds_{3,4}}$ and the overall gain of this stage is given by Eq.~\ref{eq:gain_1}~\cite{allen2011cmos}.

\begin{equation}
    A_{V1}=\frac{gm_{1,2}}{gds_{1,2}+gds_{3,4}}
    \label{eq:gain_1}
\end{equation}

After selecting a suitable value for the DC-gain ($A_{V1}$) of the first stage and assuming equal transconductance values ($gds_{1,2} = gds_{3,4}$), we determine the transconductance value ($gds_{1,2}$) using Fig.\ref{fig:pmos_gmgd_gmid}. Then, the effective width (W) is determined by finding the intersection point between $gm_{2}/I_{D2}$ and the selected length curve obtained from the previous step on the second sizing chart (Fig.~\ref{fig:pmos_vsg_gmid}) to determine $V_{GS1}$, which is used in the calculation of $(W/L)_{3,4}$.

\subsubsection*{\textbf{Gain of 2nd Stage}}

Similarly, as above, the Gain of 2nd Stage is given by Eq.~\ref{eq:gain_2}.

\begin{equation}
    A_{V2}=\frac{gm_{6}}{gds_{6}+gds_{7}}
    \label{eq:gain_2}
\end{equation}

We design the active load for the first stage (M3, M4) using the specified values of $gds_{3,4}$ and $I_{D3,4}$. Referring to Fig.~\ref{fig:nmos_vgs_gmid}, and compute the correct $(gm/ID)_{3,4}$ using $V_{GS3,4}$ and $L_{3,4}$. It is crucial to verify this calculated value before proceeding. If the computed current efficiency exceeds our assumption, indicating a lower gain, we need to reassess $(gm/ID)_{3,4}$ and recalculate $L_{3,4}$. Conversely, if the computed current efficiency falls below the assumed value, indicating a higher gain, our assumption of $(gm/ID)_{3,4}$ is appropriate. Finally, using the second sizing chart, Fig.~\ref{fig:nmos_idw_gmid}, we determine the effective width ($W_{3,4}$).

\subsubsection*{\textbf{Common Mode Rejection Ratio (CMRR)}}

It is a parameter that measures an operational amplifier's ability to reject common signals at its input terminals. Mathematically, CMRR is the ratio of differential mode gain and common mode gain, which is given as per Eq.~\ref{eq:cmrr}~\cite{allen2011cmos}.

\begin{equation}
    CMRR=\frac{A_{vd}}{A_{CM}}=\frac{gm_{1,2}}{gds_{1,2}+gds_{3,4}}.2gm_{3,4}.R_{ss}
    \label{eq:cmrr}
\end{equation}

\begin{equation}
    gds_{5} = \frac{1}{R_{ss}}
    \label{eq:gds5}
\end{equation}

To determine the aspect ratio $(W/L)_{5}$ of M5, we compute the transconductance value $gds5$ using Eq.~\ref{eq:gds5}, and assume $I_{D5} = 2*I_{D4}$. Employing a similar approach as before, we apply the methodology to calculate $(W/L)_{5}$.

\subsubsection*{\textbf{Current Ratio of M1 and M6}}

From the slew ratios of compensation capacitance and load capacitance, the current ratios can be derived as per Eq.~\ref{eq:curr_ratio16}~\cite{mahatta}.

\begin{equation}
    \frac{I_{D1}}{I_{D6}}\leq\frac{C_{C}}{2(C_{L}+C_{C})}
    \label{eq:curr_ratio16}
\end{equation}

\subsubsection*{\textbf{Phase Margin}}

Phase margin measures the amount of additional phase shift that can be introduced into the system before instability occurs. For design considerations, we have used the dominant pole concept and have derived the equation (Eq.~\ref{eq:pm_1}) for Phase Margin.

\begin{equation}
    PM^{\circ}=90^{\circ}-\tan^{-1}\left[\frac{GBW}{p2}\right]-\tan^{-1}\left[\frac{GBW}{z1}\right]
    \label{eq:pm_1}
\end{equation}

We have also defined a \textbf{Phase Margin control parameter ($\alpha$)} which is given in Eq.~\ref{eq:alpha_value}. This parameter $\alpha$ acts as a controlling knob for the Phase Margin of the overall system.

\begin{equation}
let,
    \alpha = \frac{\frac{gm1}{I_{D1}}}{\frac{gm6}{I_{D6}}}
    \label{eq:alpha_value}
\end{equation}

\begin{equation}
    PM^{\circ}=90^{\circ}-\tan^{-1}\left[\alpha\frac{I_{D1}}{I_{D6}}\frac{C_{L}}{C_{c}}\right]-\tan^{-1}\left[\alpha\frac{I_{D1}}{I_{D6}}\right]
    \label{eq:pm_2}
\end{equation}

To attain the desired phase margin as specified in Eq.~\ref{eq:pm_2}. By following a similar methodology employed to determine $(W/L)_{1}$ and $(W/L)_{2}$, we can determine the aspect ratio $(W/L)_{6}$ for M6.

\subsubsection*{\textbf{Design of 2nd Stage and Current Mirror (M8)}}

The device width ratio in the current mirror configuration, such as (M5 and M8), is equal to the current ratio. Therefore, the width of device 8 (W8) can be determined using Eq.~\ref{eq:wid8}~\cite{hesham_2}.

\begin{equation}
    \mathrm{W}8\approx\frac{2}{3}W_{5}\frac{I_{D}8}{I_{D}5}
    \label{eq:wid8}
\end{equation}

\begin{figure}[h]
    \centering
    \includegraphics[width=0.97\linewidth, height= 240pt]{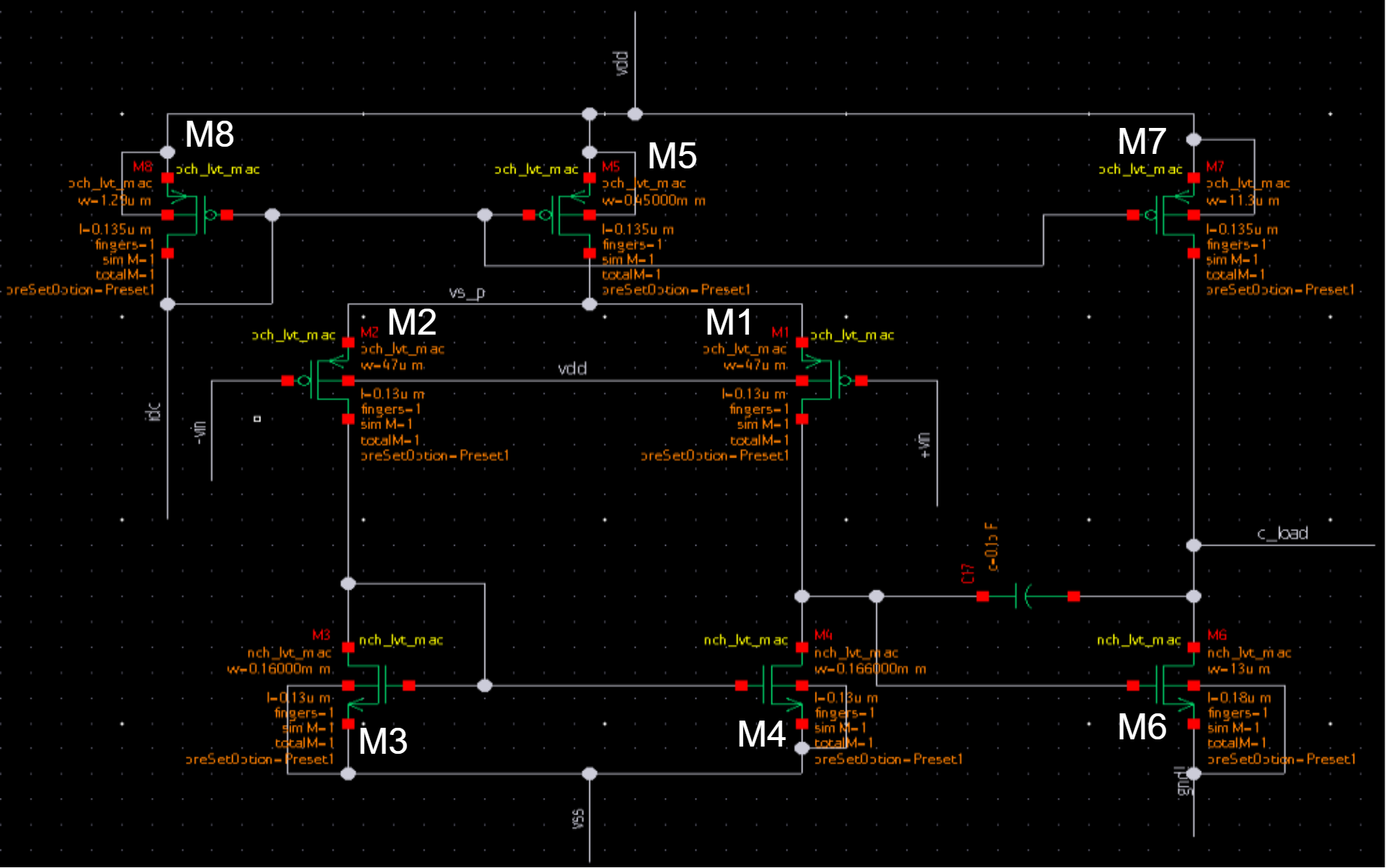}
    \caption{Internal schematic design of the differential pair}
    \label{fig:ckt_mos_level}
\end{figure}

To support the suggested process, a two-stage operational amplifier (as depicted in Fig.~\ref{fig:ckt_mos_level}) is designed using the $gm/ID$ methodology and simulated utilizing the 65 nm Low Voltage Threshold (lvt) MOSFET (tsmcN65 technology). 


\subsection{Calculated Dimensions of the MOSFETs}

After deriving all the dimensions of all the MOSFETs ($M_{1}$ to $M_{8}$), the dimensions are listed in Table.~\ref{tb:mos-specs}.

\begin{table}[h]
\centering
\caption{MOS Specifications}
\label{tb:mos-specs}
\begin{tabular}{|l|l|l|}
\hline
Mos no & W ($\mu m$)                   & L ($\mu m$)                      \\ \hline
M1     & \multirow{2}{*}{47}  & \multirow{2}{*}{0.13} \\ \cline{1-1}
M2     &                       &                        \\ \hline
M3     & \multirow{2}{*}{160} & \multirow{2}{*}{0.13} \\ \cline{1-1}
M4     &                       &                        \\ \hline
M6     & 13                   & 0.18                  \\ \hline
M5     & 450                  & \multirow{3}{*}{0.135}  \\ \cline{1-2}
M7     & 11.3                 &                        \\ \cline{1-2}
M8     & 1.29                 &                        \\ \hline
\end{tabular}

\end{table}

\section{Results and Analysis}

\subsection{AC Analysis}

\begin{figure}[h]
    \centering
    \includegraphics[width=0.85\linewidth, height=150pt]{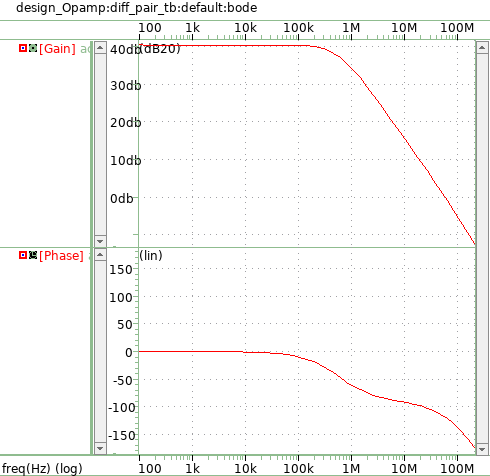}
    \caption{AC analysis of the Op-Amp}
    \label{fig:ac-anal-opamp}
\end{figure}

The design parameters for the two-stage (Miller) Op-amp depicted in Table.~\ref{tb:mos-specs} are determined through the initial iteration based on the proposed design procedure. Table.~\ref{tb:spec_opamp_calc} displays the calculated design parameters for the Op-Amp and Fig.~\ref{fig:ac-anal-opamp} shows the Bode plot of the designed Op-Amp.

\subsection{Bandwidth, Phase Margin and Power Considerations}

The performance of the designed operational amplifier (OpAmp) is evaluated in terms of bandwidth, phase margin, and power considerations. Table.~\ref{tb:spec_opamp_calc} illustrates the obtained values for these parameters. 

\begin{table}[htbp]
\centering
\caption{Performance of the Operational Amplifier (This work vs previous similar works)}
\label{tb:spec_opamp_calc}
\begin{tabular}{|l|l|l|l|} 
\hline
\textbf{Specifications}             & \textbf{This work} & \textbf{Work~\cite{singh_2019}} & \textbf{Work~\cite{journal_2018_ieee}}  \\ 
\hline
Technology              & 65nm & 180nm & 500nm     \\ 
\hline
Supply Voltage ($V_{dd}$)              & 0.9V & 1.8V & 3.3V  \\ 
\hline
IRNV~\footnote{Input Referred Noise Voltage } ($S_{n}(f)$)  & 8 nV/$\sqrt{\text{Hz}}$ & - & - \\ 
\hline
Load Capacitance ($C_{L}$)                & 4 pF & - & - \\ 
\hline
Gain Bandwidth Product (GBW)   & 60 MHz & - & - \\ 
\hline
DC gain ($A_o$)                  & 40.4 dB  & 65.88 dB & 32 dB \\ 
\hline
Phase Margin (in degrees)                  & 61.3$^\circ$ & - & -    \\ 
\hline
Common Mode Input (low)       & $\leq$ 0.125 V & - & -  \\ 
\hline
Common Mode Input (high)     & $\geq$ 5V & - & -     \\ 
\hline
CMRR                         & 68 dB & 136.05 dB  & 88 dB    \\ 
\hline
Slew Rate                    & 18 V/$\mu$s & - & -  \\ 
\hline
Power Dissipation                   & 0.29 mW & 1.3 mW & 0.28mW  \\ 
\hline
\end{tabular}
\end{table}

\section{Conclusion}

In this work, we have developed a low-power, high-gain operational amplifier (op-amp) specifically designed for biomedical instruments. Our op-amp boasts a DC gain of 40.4 dB, a gain bandwidth product of 60 MHz, and a power dissipation of only 0.29 mW. We conducted a thorough comparison with the work of Singh et al.~\cite{singh_2019} and found that our design significantly reduces power requirements while achieving a significantly higher gain within a similar power range, surpassing the results reported in the work by Oreggion et al.~\cite{journal_2018_ieee}.




\newpage
\bibliographystyle{ieeetr}
\bibliography{citation}
\vspace{12pt}

\end{document}